\documentclass[journal]{IEEEtran}
\usepackage{multirow}
\usepackage[dvipdfmx]{graphicx} 
\usepackage{epsfig} 
\usepackage{mathptmx} 
\usepackage{times} 
\usepackage{amsmath} 
\usepackage{amssymb}  
\usepackage{subfigure}
\usepackage{textcomp,gensymb}
\usepackage{url}
\usepackage{xcolor}

\ifCLASSINFOpdf
\else
\fi
\begin{document}
%
\title{Co-Optimization of On-Ramp Merging and Plug-In Hybrid Electric Vehicle Power Split Using Deep Reinforcement Learning}
%

\author{Yuan Lin,~\IEEEmembership{Member, IEEE}, John McPhee,
        and Nasser L. Azad,~\IEEEmembership{Member, IEEE}
\thanks{Yuan Lin is with the Shien-Ming Wu School of Intelligent Engineering at the South China University of Technology, Guangzhou, China 511442. {\tt\small yuanlin@scut.edu.cn}}%
\thanks{John McPhee and Nasser L. Azad are with the Systems Design Engineering Department at the University of Waterloo, Ontario, Canada N2L 3G1. {\tt\small mcphee@uwaterloo.ca; nlashgarianazad@uwaterloo.ca}}%
}

\markboth{}
{Shell \MakeLowercase{\textit{et al.}}: Bare Demo of IEEEtran.cls for IEEE Journals}
%



\maketitle

\begin{abstract}

Current research on Deep Reinforcement Learning (DRL) for automated on-ramp merging neglects vehicle powertrain and dynamics. This work considers automated on-ramp merging for a power-split Plug-In Hybrid Electric Vehicle (PHEV), the 2015 Toyota Prius Plug-In, using DRL. The on-ramp merging control and the PHEV energy management are co-optimized such that the DRL policy directly outputs the power split between the engine and the electric motor. The testing results show that DRL can be successfully used for co-optimization, leading to collision-free on-ramp merging. When compared with sequential approaches wherein the upper-level on-ramp merging control and the lower-level PHEV energy management are performed independently and in sequence, we found that co-optimization results in economic but jerky on-ramp merging while sequential approaches may result in collisions due to neglecting powertrain power limit constraints in designing the upper-level on-ramp merging controller.

\end{abstract}


%
\IEEEpeerreviewmaketitle

\section{Introduction}

Automated on-ramp merging is an on-going research topic that has seen different methods proposed \cite{rios2016survey}. The early approaches include heuristics-based methods such as Intelligent-Driver-Model (IDM) \cite{treiber2000congested} and optimization-based methods such as Model Predictive Control (MPC) \cite{rios2016survey}. The solution framework can be centralized wherein merging is controlled by a roadside unit to some extent \cite{rios2017automated}, or decentralized wherein the merging vehicle merges on ramp without the control by a roadside unit \cite{tran2019model}.

Recently, reinforcement learning (RL), particularly DRL, has been used to solve the automated on-ramp merging problem \cite{nishi2019merging}. In RL, an agent learns to maximize cumulative discounted reward for optimal decision-making and control in an environment \cite{sutton2018reinforcement}. DRL that utilizes deep (multi-layer) neural nets as the policy has seen breakthroughs as its trained policy surpassed human champions in playing complex board games \cite{silver2016mastering}. It has been demonstrated that DRL offers near-optimal control performance and high modeling-error tolerance while requiring much less computation time than MPC \cite{lin2020comparison}.

Currently DRL-based on-ramp merging is mostly decentralized. A popular approach is integrated decision and control such that DRL outputs acceleration whose values reflect the decision to merge behind or ahead a vehicle on the main road \cite{hu2019interaction,bouton2019cooperation}. It is shown that DRL could lead to collision-free on-ramp merging \cite{lin2020anti}. However, current DRL-based on-ramp merging only considers the merging vehicle as a point-mass kinematic model, neglecting vehicle powertrain and dynamics. As far as our work is concerned, current literature has not considered DRL-based on-ramp merging for a PHEV.

In recent years, hybrid electric vehicles (HEVs) are becoming popular among consumers as they offer better fuel economy compared to conventional internal combustion engine vehicles \cite{liu2008modeling}. HEVs have two power sources: the engine and the battery. HEVs can save fuel via regenerative braking and engine stops in stop-and-go traffic \cite{gao1999investigation,shao2019optimal}. Compared to HEVs, PHEVs have a larger battery that can be plugged into an external power grid \cite{taghavipour2019intelligent,jeong2019analysis}. Thus, PHEVs could provide an initial electric drive, similar to battery electric vehicles. HEVs and PHEVs offer a transition towards the pure electric vehicle era.

The energy management for HEVs or PHEVs is a challenging problem that has spurred abundant research \cite{taghavipour2015real,taghavipour2016comparative}. The goal of energy management is to minimize energy consumption by deciding the power split between the engine and the motor. There are three main approaches to energy management: rule-based, optimization-based, and RL-based. Rule-based methods utilize the battery state of charge (SOC) and the power demand to define the power split rules \cite{lin2003power,banvait2009rule}. For HEVs, a typical rule-based method is Charge Sustaining (CS) wherein the battery SOC is sustained within a certain range; for PHEVs, an initial Charge Depleting (CD) is added before CS to deplete the battery SOC first. Optimization-based energy management utilizes trip information and vehicle powertrain and dynamics modeling to solve for the optimal power split \cite{kim2010optimal,musardo2005ecms,borhan2011mpc,serrao2011comparative}.

Recently, RL has emerged as an alternative to classical optimization for HEV or PHEV energy management \cite{lin2014reinforcement,liu2014power}. With vehicle powertrain and dynamics modeling, RL-based energy management utilizes pre-defined drive cycles to train optimal power-split policies \cite{li2019deep,xu2020ensemble,qi2019deep,liu2019heuristic}. Studies show that RL could provide near-optimal energy management compared to benchmark dynamic programming solutions \cite{liu2017reinforcement,zhu2020energy}. However, there is currently no literature addressing RL-based HEV or PHEV energy management during on-ramp merging.

To handle on-ramp merging for a PHEV, both co-optimization and sequential approaches can be used. In co-optimization, the upper-level vehicle behavior such as on-ramp merging and the lower-level PHEV energy management are simultaneously optimized. In sequential approaches, the upper-level vehicle behavior and the lower-level PHEV energy management are performed independently and in sequence. Co-optimization is shown to result in better fuel economy but higher jerk for cruise and car-following behaviors \cite{chen2019series,chen2020iterative,li2017fuel}. There is recent work on co-optimization for centralized on-ramp merging \cite{xu2021decentralized}. However, current work on co-optimization has only considered classical optimization to solve for the optimal solutions. To the best of our knowledge, there is no published work on HEV or PHEV co-optimization using DRL.

\begin{figure}[!htbp]
\centering
\includegraphics[width=3.4in]{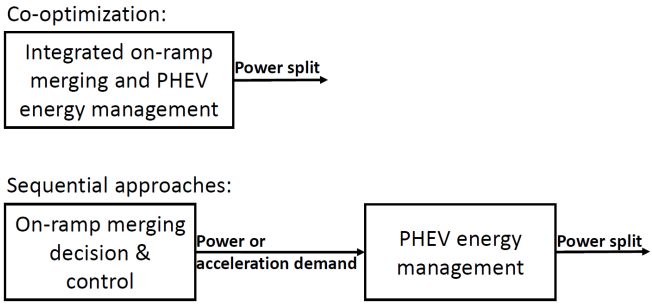}
\caption{Co-optimization versus sequential approaches.}
\label{fig:co_sequential}
\end{figure}

This work focuses on PHEV co-optimization using DRL for on-ramp merging. The co-optimization dictates that the upper-level on-ramp merging and the lower-level PHEV energy management are simultaneously optimized via a single DRL framework such that the DRL policy directly outputs the power split. For sequential approaches, the upper-level DRL-based merging controller outputs the power or acceleration demand, which is then given to the lower-level rule-based PHEV energy management to decide the power split, see Fig.~\ref{fig:co_sequential}.

The contributions of this work include:

(1) Co-optimization of the upper-level on-ramp merging and the lower-level PHEV energy management using RL;

(2) Comparison of co-optimization and sequential approaches for PHEV on-ramp merging;

(3) RL-based on-ramp merging with vehicle powertrain and dynamics considered.

This work allows us to evaluate if PHEV co-optimization using DRL is workable. It also enables us to observe the PHEV powertrain response during DRL-based on-ramp merging.

The rest of the paper is organized as follows: Section II presents the Control-Oriented Model (COM) of the PHEV, the 2015 Toyota Prius Plug-In. Section III illustrates the on-ramp merging environment in a driving simulator. Section IV explains the DRL framework for co-optimization. Section V presents the sequential approaches. Section VI shows the DRL training and testing results. Section VII draws conclusions and suggests future work.

\section{PHEV Control-Oriented Model}

The PHEV COM is a simplified model of the PHEV that includes the powertrain configuration, battery, and vehicle dynamics modeling. The PHEV COM is established for studying the PHEV powertrain response and energy consumption during on-ramp merging.

\subsection{PHEV powertrain}

\begin{figure}[!htbp]
\centering
\includegraphics[width=3in]{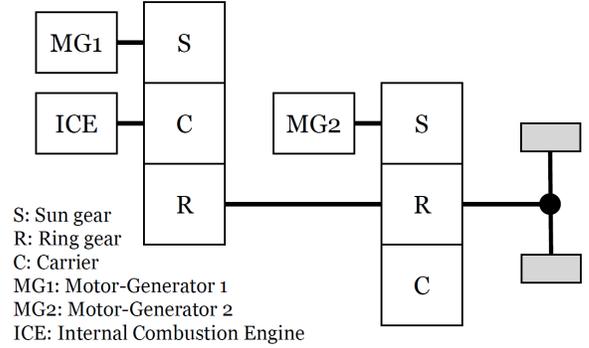}
\caption{Powertrain configuration of the 2015 Toyota Prius Plug-In.}
\label{fig:powertrain}
\end{figure}

Fig.~\ref{fig:powertrain} shows the powertrain configuration of the 2015 Toyota Prius Plug-In. The PHEV has an engine, a battery, two planetary gears, and two motor-generators: MG1 and MG2. MG1 is mostly used as a generator to aid the engine to work in its optimal operation while MG2 is used as both a motor to propel the vehicle and a generator for regenerative braking. As the actual powertrain control algorithms are unknown, we base the COM on the powers of components: the engine power $P_{eng}$, motor-generator combined power $P_{mg}$, friction brake power $P_{fbk}$, and power demand before transmission loss $P_d$. These powers are related by the following equation.
\begin{equation}
\begin{split}
P_d = P_{eng} + P_{mg} + P_{fbk}\\
\end{split}
\label{eq:Pd0}
\end{equation}

The battery power $P_b$ is considered to be related to the motor-generator combined power $P_{mg}$ as follows.
\begin{equation}
\begin{split}
P_b = P_{mg}/\eta_m + P_{aux}, P_{mg} \geq 0\\
P_b = P_{mg}\eta_g + P_{aux}, P_{mg} < 0
\end{split}
\label{eq:Pmg}
\end{equation}
where $\eta_m$ is the motor efficiency when the combined effect of motor-generator is propelling the vehicle, $\eta_g$ is the generator efficiency when the combined effect of motor-generator is storing energy, and $P_{aux}$ is the auxiliary power needed for vehicle accessories which don't include air conditioning.

The engine fuel rate $\dot{m_f}$ is considered to have a linear relationship with the engine power $P_{eng}$, assuming that the engine always runs in its optimal operation.
\begin{align}
\begin{split}
\dot{m}_f = a_1 P_{eng} + a_2, P_{eng} > 0\\
\dot{m}_f = 0, P_{eng} = 0
\end{split}
\label{eq:fuel_rate}
\end{align}
where $a_1$ and $a_2$ are coefficients.

\subsection{Battery model}

The battery SOC is defined as $\dot{SOC} = - I_b/Q_{max}$ where $I_b$ is the battery current and $Q_{max}$ is the battery capacity. The battery is modeled as a simple circuit system with the battery power computed as $P_b = I_b V_{oc} - I_b^2 R_b$ where $V_{oc}$ is the battery open circuit voltage and $R_b$ is the battery internal resistance. Thus, the battery $SOC$ in terms of the battery power $P_b$ is
\begin{align}
\begin{split}
\dot{SOC} = - \frac{V_{oc} - \sqrt{V_{oc}^2 - 4 R_b P_b}}{2 R_b Q_{max}}\\
\end{split}
\label{eq:SOC}
\end{align}
where $V_{oc}$ and $R_b$ are considered functions of the SOC \cite{liu2008modeling}. Quadratic functions are used to estimate their relationships.
\begin{align}
\begin{split}
V_{oc} = b_1 SOC^2 + b_2 SOC + b_3\\
R_b = c_1 SOC^2 + c_2 SOC + c_3
\end{split}
\label{eq:Voc_Rb}
\end{align}
where $b_1$, $b_2$, $b_3$, $c_1$, $c_2$ and $c_3$ are coefficients. Eqns.~\ref{eq:SOC} and ~\ref{eq:Voc_Rb} determine the battery SOC dynamics once $P_b$ is known.

\subsection{Vehicle dynamics}

The power demand before transmission loss $P_d$ achieves the vehicle acceleration by overcoming resistant forces, which include the aerodynamic drag, the rolling resistance, and the gravity due to road grades.

\begin{equation}
\begin{split}
ma = \frac{P_d\eta_t^k}{v} - [0.5 C_d \rho A v^2 + (\mu + \mu_2 v) mg cos(\theta) + mg sin(\theta)]\\
\dot{v} = a\\
k = 1, P_d \geq 0\\
k = -1, P_d < 0
\end{split}
\label{eq:Pd}
\end{equation}
where $\eta_t \in$(0,1) represents the transmission efficiency and the superscript $k$ is either 1 or -1 for non-negative or negative power demand, respectively. When $P_d \geq 0$, $\eta_t^k = \eta_t$, which means that the non-negative power demand propels the vehicle with transmission loss. When $P_d < 0$, $\eta_t^k = \eta_t^{-1}$, which means that the negative power demand results from vehicle braking with transmission loss. In Eqn.~\ref{eq:Pd}, $m$ is the vehicle mass, $a$ is the acceleration, $v$ is the velocity, $C_d$ is the aerodynamic drag coefficient, $A$ is the vehicle frontal area, $\mu$ and $\mu_2$ are rolling coefficients, $g$ is the gravity constant, and $\theta$ is the road grade in radians. In this work, we only consider flat roads with $\theta=0$.

\subsection{Energy cost}

The energy cost of a PHEV is the combination of both the fuel and electricity monetary costs \cite{vajedi2015ecological}. The instantaneous energy cost during a time step $\Delta t$ is computed as
\begin{equation}
\begin{split}
c = (k_f \dot{m_f} + k_e \frac{P_b}{\eta_b \eta_{chr}})\Delta t
\end{split}
\label{eq:cost}
\end{equation}
where $k_f$ is the unit price of fuel in US dollars (USD), $\dot{m_f}$ is the fuel rate, $k_e$ is the unit price of electricity power in USD, $\eta_b$ is the battery efficiency, $\eta_{chr}$ is the external charger efficiency. The unit prices $k_f=0.93$ USD/kg and $k_e=0.13$ USD/kWh are the US national averages in 2019.

The constraints of the vehicle variables are
\begin{equation}
\begin{split}
0 \leq P_{eng} \leq P_{eng,max}\\
P_{brk,min} \leq P_{brk} \leq 0\\
P_{g,min} \leq P_{mg} \leq P_{m,max}\\
P_{b,min} \leq P_b \leq P_{b,max}
\end{split}
\label{eq:constraints}
\end{equation}
where $P_{eng,max}$ is the maximum engine power, $P_{brk,min}$ is the minimum friction brake power, $P_{g,min}$ is the minimum generator power, $P_{m,max}$ is the maximum motor power, and $P_{b,min}$ and $P_{b,max}$ are the minimum and maximum battery powers, respectively. The motor-generator power is constrained by not only its own limits but also the battery power limits. All the COM parameter values are determined according to a high-fidelity vehicle model that has been validated against a 2015 Toyota Prius Plug-In in our previous work \cite{taghavipour2013high,buggaveeti2017longitudinal}.


\section{On-Ramp Merging Environment}

\begin{figure}[!htbp]
\centering
\includegraphics[width=3.2in]{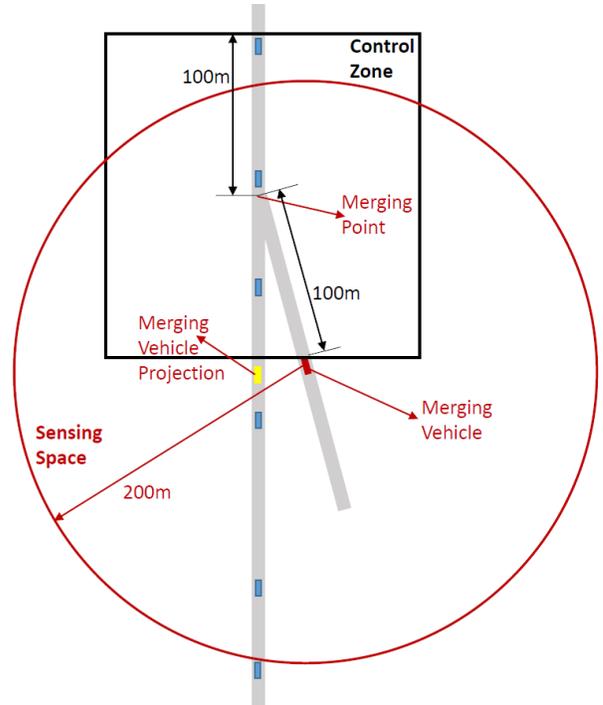}
\caption{Schematic for on-ramp merging.}
\label{fig:schematic}
\end{figure}

The merging environment is created in the Simulation of Urban Mobility (SUMO) driving simulator for DRL training and testing, see Fig.~\ref{fig:schematic}. We consider a single-lane main road and a single-lane on-ramp. The main road has a speed limit of $v_{limit} = 29.06$m/s (65mph). There are intelligent vehicles on the main road that perform car following and collision avoidance based on the IDM. Each of the intelligent vehicles has a desired speed of $\alpha \, v_{limit}$ where $\alpha$ is a constant. $\alpha$ is based on a Gaussian distribution with mean 1 and standard deviation 0.1 with values clipped within [0.85, 1.15]. Thus, the desired speeds of the intelligent vehicles are within [24.70,33.42]m/s (55-75mph). The intelligent vehicles are generated at the bottom of the main road with a probability of 0.5 per second. The different desired speeds, the probabilistic traffic generation, and the IDM parameter variances result in significantly different traffic densities at the merging junction at different time instants. The normal acceleration of the intelligent vehicles lies within [$a_{min}, a_{max}$] = [-4.5, 2.6]m/s$^2$. However, during emergency braking when the merging vehicle is too close, the acceleration can further decrease to -9m/s$^2$.

In this work, the merging vehicle is only controlled in a control zone defined as 100m behind the merging point and 100m ahead of the merging point. The merging point is defined as the intersection between the main road and the on-ramp. We assume that there are no other vehicles in front of the merging vehicle on the ramp. The merging vehicle has a sensing radius of 200m which could be enabled by advanced perception sensors \cite{fersch2017challenges,hecht2018lidar}. The merging vehicle has no wireless communication and there is no centralized control or coordination.

The vehicles on the main road are designed to have the right of way and only interact with the merging vehicle when the merging vehicle enters the junction between the main road and the on-ramp. The junction is created by the SUMO driving simulator and covers the road blocks around 15m away from the merging point. When the merging vehicle enters the junction on the ramp, its projection on the main road is used by the main-road vehicles to perform car following based on the IDM. Its projection on the main road is defined to have the same distance to the merging point as it does on the ramp.

For every merging episode, the merging vehicle is generated 100m behind the merging point on the ramp (at the control zone bottom) with an initial speed randomly distributed in [22.35, 26.82]m/s (50-60mph). This is based on the recommendation by the US Department of Transportation that the merging vehicle reaches at least 50mph to merge onto a highway of 65mph speed limit \cite{aashto2001policy}. The initial acceleration of the merging vehicle is zero. The initial battery SOC of the merging vehicle is randomly distributed in [0.3,0.9]. A merging episode terminates when the merging vehicle stops, collides with another vehicle, or reaches the destination which is 100m ahead of the merging point.

\section{DRL Framework for Co-Optimization}

To provide the training dataset for DRL, the DRL environment state, the action for the merging vehicle, and the reward function are defined. In this section, we introduce the DRL framework for co-optimization.

\subsection{Environment state}

The DRL environment state $S$ consists of the states of 5 vehicles: the merging vehicle or its projection, along with 2 preceding and 2 following vehicles. The projection of the merging vehicle has the same distance to the merging point as the merging vehicle, see Fig.~\ref{fig:schematic}. For the 2 preceding and 2 following vehicles, the states are their distances to the merging point and velocities. For the merging vehicle, the state includes its distance to the merging point $d$, velocity $v$, and acceleration $a$. The inclusion of the merging vehicle's velocity enables the calculation of its acceleration when the power demand is given, see Eqn.~\ref{eq:Pd}. The inclusion of acceleration enables the calculation of jerk since there is a DRL reward that penalizes high jerk, see the following Reward subsection. The inclusion of the battery SOC enables finding its relationship with the battery power, see Eqn.~\ref{eq:SOC}. Thus, the DRL environment state $S$ is
\begin{equation}
\begin{split}
S = [d_{p2}, v_{p2}, d_{p1}, v_{p1}, d, v, a, SOC, d_{f1}, v_{f1}, d_{f2}, v_{f2}]
\end{split}
\label{eq:state}
\end{equation}
where $p2$ denotes the second preceding vehicle, $p1$ denotes the first preceding vehicle, $f1$ denotes the first following vehicle, and $f2$ denotes the second following vehicle.

When there are fewer than 2 preceding or 2 following vehicles in the circular sensing range of the merging vehicle, virtual vehicles are introduced to deliberately construct the five-vehicle environment state $S$. The introduction of virtual vehicles is needed since DRL requires a fixed-size environment state. The virtual vehicles are positioned at the intersections between the merging vehicle's circular sensing space and the main lane, with velocities as the speed limit of the main lane and zero acceleration.

\subsection{Action}

For co-optimization, DRL directly outputs the power split. Therefore, the DRL action $A$ includes the engine power $P_{eng}$, and the combined power $P_{cb}$ of the motor-generator and friction brake powers. That is,
\begin{equation}
\begin{split}
A = [P_{eng}, P_{cb}]\\
P_{cb} = P_{mg} + P_{fbk}
\end{split}
\label{eq:action}
\end{equation}
When the combined power $P_{cb}$ is positive, we set $P_{fbk}=0$; then $P_{mg} = P_{cb}$, meaning that the motor propels the vehicle. When the combined power $P_{cb}$ is negative, the vehicle brakes. During braking, the generator is utilized first to store the braking energy into the battery; the friction brake is only used when the braking power exceeds the generator power limit. With this rule for the combined power $P_{cb}$, we guarantee that regenerative braking is prioritized over friction brake for energy-saving. In contrast, if the motor-generator $P_{mg}$ and friction brake $P_{fbk}$ powers are outputted separately by DRL, it adds to the challenge for DRL to find the policy that perfectly prioritizes regenerative braking.

\subsection{Reward}

Several rewards are designed at each time step for the desired merging performance:

(1) The merging vehicle is encouraged to merge midway between two vehicles with the velocity of the first preceding vehicle $v_{p1}$. We do not require the merging vehicle's velocity to be dependent on that of the first following vehicle because otherwise the braking of the first following vehicle leads to the braking of the merging vehicle. We define the following penalizing reward after the merging vehicle merges onto the main road:
\begin{equation}
\begin{split}
r_m = -w_m (\lambda + \frac{|v_{p1} - v|}{\Delta v_{max}})
\end{split}
\label{eq:r_m}
\end{equation}
where $w_m$ is the weight, $\Delta v_{max}=5$m/s is the maximum allowed velocity difference between the first preceding and the merging vehicles, and $\lambda$ is the midway ratio defined as follows:
\begin{equation}
\begin{split}
\lambda = \frac{|dd_{p1} - dd_{f1}|}{dd_{p1} + dd_{f1}}
\end{split}
\label{eq:midway}
\end{equation}
where $dd_{p1}$ is the inter-vehicular distance between the first preceding and the merging vehicles, and $dd_{f1}$ is the inter-vehicular distance between the merging and the first following vehicles. When $dd_{p1}=dd_{f1}$, $\lambda=0$ and the merging vehicle merges exactly midway. When $dd_{p1}=0$ or $dd_{f1}=0$, $\lambda=1$ and the merging vehicle has zero distance to either the first preceding or the first following vehicle.

(2) Merging may result in congestion on the main road. Thus, we define a penalizing reward to reduce the braking of the first following vehicle.
\begin{equation}
\begin{split}
r_b = -w_b \frac{|a_{f1}|} {max(|a_{min}|,a_{max})}, a_{f1}<0
\end{split}
\label{eq:r_b}
\end{equation}
where $w_b$ is the weight and $a_{f1}$ is the acceleration of the first following vehicle.

(3) For passenger comfort, we define a penalizing reward to reduce the jerk of the merging vehicle $j$.
\begin{equation}
\begin{split}
r_j = -w_j \frac {|j| - j_0} {j_{max} - j_0}, |j| > j_0\\
j = \dot{a}\\
j_{max} = \frac{a_{max} - a_{min}}{\Delta t}
\end{split}
\label{eq:r_j}
\end{equation}
where $w_j$ is the weight, $\Delta t$=0.1s is the time step, and $j_0$=3m/s$^3$ is the maximum allowed absolute jerk value within which the passenger feels comfortable \cite{batra2019real}.

(4) For economic merging, we define a penalizing reward to reduce the instantaneous energy cost of the merging vehicle $c$.
\begin{equation}
\begin{split}
r_c = -w_c \frac{c}{c_{max}}
\end{split}
\label{eq:r_c}
\end{equation}
where $w_c$ is the weight and $c_{max}$ is the maximum energy cost at each time step. $c_{max}$ is computed based on the instantaneous energy cost equation Eqn.~\ref{eq:cost} when the engine and battery powers are set at maximum.

(5) A penalizing reward of $r_{stop}=$-1 is given when the merging vehicle stops and the episode terminates.

(6) A penalizing reward of $r_{collision}=$-1 is given when the merging vehicle collides with any vehicles and the episode terminates. Collision is defined when the inter-vehicular distance is smaller than 2.5m.

(7) A reward of $r_{success}=$1 is given when the merging vehicle reaches the destination which is 100m ahead of the merging point and the episode terminates.

Note that there is no weighting for the terminating rewards (5) to (7) above. The tuning of the weights for rewards (1) to (4) is based on trial and error. The weight tuning priority order is safety+traffic free flow, passenger comfort, and energy saving. Reward (1) demands merging midway, which contributes to safety. Reward (2) penalizes the braking of the first flowing vehicle for traffic free flow, which also contributes to safety between the first following and the merging vehicles. Thus, rewards (1) and (2) are tuned together for safety+traffic free flow. Particularly, we first set the weights for jerk and energy cost to zeros and tune the weights for rewards (1) and (2) for zero collisions and zero stops. Then we increase the weight for jerk to maximum while maintaining zero collisions and zero stops. Last, we increase the weight for energy cost to maximum while maintaining zero collisions, zero stops, and average jerk less than or equal to 1m/s$^3$. Table~\ref{table:weights} shows the reward weights.

\begin{table}[h]
\caption{Reward weights.}
\begin{center}
\begin{tabular}{|p{6.8cm}|c|}
\hline
Weight for merging midway with velocity of first preceding vehicle $w_m$ & 0.015\\
\hline
Weight for penalizing braking of first following vehicle $w_b$ & 0.015\\
\hline
Weight for penalizing jerk $w_j$ & 0.1\\
\hline
Weight for penalizing energy cost $w_c$ & 0.005\\
\hline
\end{tabular}
\end{center}
\label{table:weights}
\end{table}


\section{Sequential Approaches}

In order to evaluate co-optimization, two sequential approaches are considered for comparison purposes. In the first sequential approach (sequential approach I), the upper-level DRL-based merging controller outputs the power demand, which is then given to the lower-level PHEV blended CD energy management to decide the power split. In the second sequential approach (sequential approach II), the upper-level DRL-based merging controller outputs the acceleration demand, which is then given to the lower-level PHEV blended CD energy management to decide the power split.

Since most published work utilizes acceleration demand as the output of the upper-level controller, sequential approach II may be better received in the current literature. Moreover, sequential approach II does not require powertrain information to design the upper-level merging controller, which makes it more general than sequential approach I and co-optimization.

For sequential approach I, since the upper-level DRL-based merging controller outputs the power demand, the DRL action is the power demand.
\begin{equation}
\begin{split}
A_1 = P_d
\end{split}
\label{eq:action1}
\end{equation}

The rewards and their weights are the same as those for co-optimization except for the reward to penalize energy cost, which is defined as
\begin{equation}
\begin{split}
r_{c1} = -w_c \frac{P_d}{max(|P_{g,min}+P_{brk,min}|, P_{m,max}+P_{eng,max})}
\end{split}
\label{eq:r_c1}
\end{equation}
The weight $w_c$ is the same as that for co-optimization.

For sequential approach II, since the upper-level DRL-based merging controller outputs the acceleration demand, the DRL action is the acceleration demand.
\begin{equation}
\begin{split}
A_2 = a_d\\
a_{min} \leq a_d \leq a_{max}
\end{split}
\label{eq:action2}
\end{equation}

The rewards and their weights are the same as those for co-optimization except for the reward to penalize energy cost, which is defined as
\begin{equation}
\begin{split}
r_{c2} = -w_c \frac{a_d}{max(|a_{min}|,a_{max})}
\end{split}
\label{eq:r_c2}
\end{equation}
The weight $w_c$ is the same as that for co-optimization.

For both sequential approaches, the upper-level DRL training does not involve PHEV energy management. Thus, compared to co-optimization, the DRL environment state for both sequential approaches does not include the battery SOC.
\begin{equation}
\begin{split}
S_{12} = [d_{p2}, v_{p2}, d_{p1}, v_{p1}, d, v, a, d_{f1}, v_{f1}, d_{f2}, v_{f2}]
\end{split}
\label{eq:state12}
\end{equation}

For both sequential approaches, the lower-level PHEV energy management method is the blended CD mode. In blended CD, if the power demand can be supplied by the battery only, the engine is off and the vehicle is in electric drive mode. Otherwise, the engine is on and supplies all the power demand; if the engine cannot supply all the power demand, the battery makes up for the power insufficiency.

The blended CD mode prioritizes electric drive over engine power, which results in low energy costs as the cost of electricity is lower than that of fuel. We do not consider the CS mode since CS sometimes prioritizes engine power when the power demand is above a threshold, which results in higher energy costs. The blended CD mode allows us to evaluate if co-optimization could offer comparably low energy costs.

In our simulation, merging usually takes less than 10 seconds, so the resulting SOC change is very small. There is no concern of constraining the SOC above a threshold during the short-period merging for the sequential and co-optimization approaches.

\section{DRL Training and Testing}

\subsection{The DRL algorithm}
RL can be formulated as a Markov Decision Process. That is, at time step $t$, an RL agent observes the environment state $s_t$, performs an action $a_t$ based on its policy, and subsequently receives a reward $r_t$ as the environment moves to the next state $s_{t+1}$. The goal of RL is to learn a policy that maximizes the cumulative discounted reward $\sum_{t=0}^{t=T} \gamma^{\,t} r_t$ where $T$ is the total time step and $\gamma$ is the discount factor.

The DRL algorithm used here is Soft Actor-Critic (SAC) \cite{haarnoja2018soft}. SAC is an entropy-regulated actor-critic algorithm wherein the RL agent is given an additional reward related to the entropy of the policy. The actor is the policy $\pi_{\phi}$ with the deep neural net parameters $\phi$ and the critic is the Q-value $Q_{\theta}$ with the deep neural net parameters $\theta$. Q-value is defined as the cumulative discounted reward from the current time step $t$ to the total time step $T$: $Q_{\theta}(s_t,a_t) = \sum_{\tau = t}^{\tau = T} \gamma^{\,\tau - t} r_{\tau}$ where $\tau$ denotes a time step between $t$ and $T$.

The critic network $\theta$ is trained according to Bellman's optimality principle by minimizing the root-mean-squared loss $L_t = \{r_t + \gamma[Q_{\theta}(s_{t+1},\pi_{\phi}(s_{t+1})) - \alpha log \pi_{\phi}(s_{t+1})] - Q_{\theta}(s_t,a_t)\}^2$ where $\alpha$ is the weight for the entropy reward that is the log value of the policy. The policy network $\phi$ is trained by maximizing $[Q_{\theta}(s_{t},\pi_{\phi}(s_{t})) - \alpha log \pi_{\phi}(s_{t})]$. Techniques including target networks, mini-batch gradient descent/ascent, and batch normalization \cite{ioffe2015batch} are used to improve training stability and convergence. Additionally, Double Q-Learning is also used to reduce Q-value over-estimation \cite{van2016deep}.

The entropy regularization in SAC results in stochasticity and exploration, which leads to better optimum seeking and training stability. Application of SAC to complex tasks shows that SAC is better than some of the well-known DRL algorithms such as Deep Deterministic Policy Gradient (DDPG) \cite{lillicrap2015continuous} and Proximal Policy Optimization \cite{schulman2017proximal}. For more details, readers are encouraged to read the original SAC paper \cite{haarnoja2018soft}.

\subsection{DRL training}
For the co-optimization and two sequential approaches, the merging environment setting is the same. The DRL training is based on episodes. Each episode starts as the merging vehicle is generated 100m behind the merging point on the ramp, and terminates at 100m ahead of the merging point if merging is successful, or in the control zone when there are stops or collisions. The total training time is 1 million time steps for each approach, which result in roughly 12500 to 13000 episodes. Before each episode starts, there is a 10-second buffer for traffic initialization on the main road.

For SAC training for each approach, both the actor and critic networks have 2 hidden layers with 64 neurons for each hidden layer. On a desktop computer with a 16-core (32-thread) AMD processor and a GeForce RTX 2080Ti GPU, the training took around 6 hours.

\begin{figure}[htbp]
\centering
\includegraphics[width=3in]{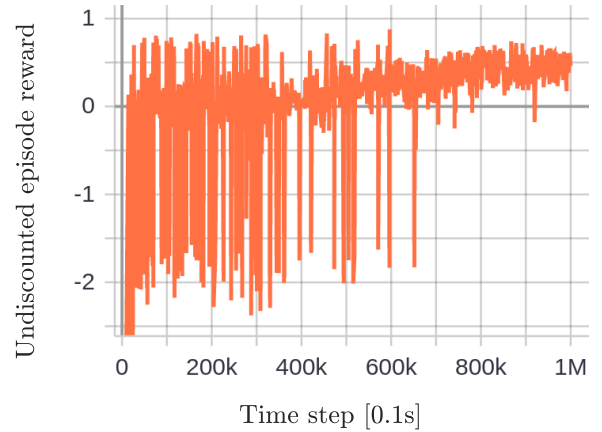}
\caption{Undiscounted episodic reward during SAC training for co-optimization.}
\label{fig:reward}
\end{figure}

The DRL training was not robust since training the same approach could lead to noticeably different testing results. Thus, we trained each approach twice and chose the trained policy that resulted in zero collisions, zero stops, and better values of the evaluation metrics defined in the following DRL testing section. Fig.~\ref{fig:reward} shows the undiscounted episodic reward during training for the co-optimization approach. The undiscounted episodic rewards for the two sequential approaches look similar and are not plotted here. From Fig.~\ref{fig:reward}, we see that the undiscounted episodic reward converges during training.

\subsection{DRL testing}

Testing is performed on the same merging environment used for training. In other words, the trained policies are not tested in unseen environments. However, each episode has a unique initial condition, which means that each episode is not the same. For each approach, the testing time is also 1 million time steps, which took around 4 hours on the same computer used for training.

In the following, we first present two testing episodes to showcase the merging dynamics and the PHEV powertrain response; then we summarize the testing results to compare the co-optimization and the two sequential approaches. The two testing episodes presented include: (1) A testing episode for co-optimization. In this episode, the merging vehicle maintains its relative positioning order between the first preceding and first following vehicles during the entire merging process, i.e., merges ahead; (2) A testing episode for sequential approach II. In this episode, the merging vehicle merges behind the first following vehicle. In addition, this episode presents the issue of power limit saturation in which the acceleration demand is not achievable due to the PHEV engine and battery power limits.

\subsubsection{A testing episode for co-optimization}

\begin{figure*}[htbp]
\centering
\includegraphics[width=7in]{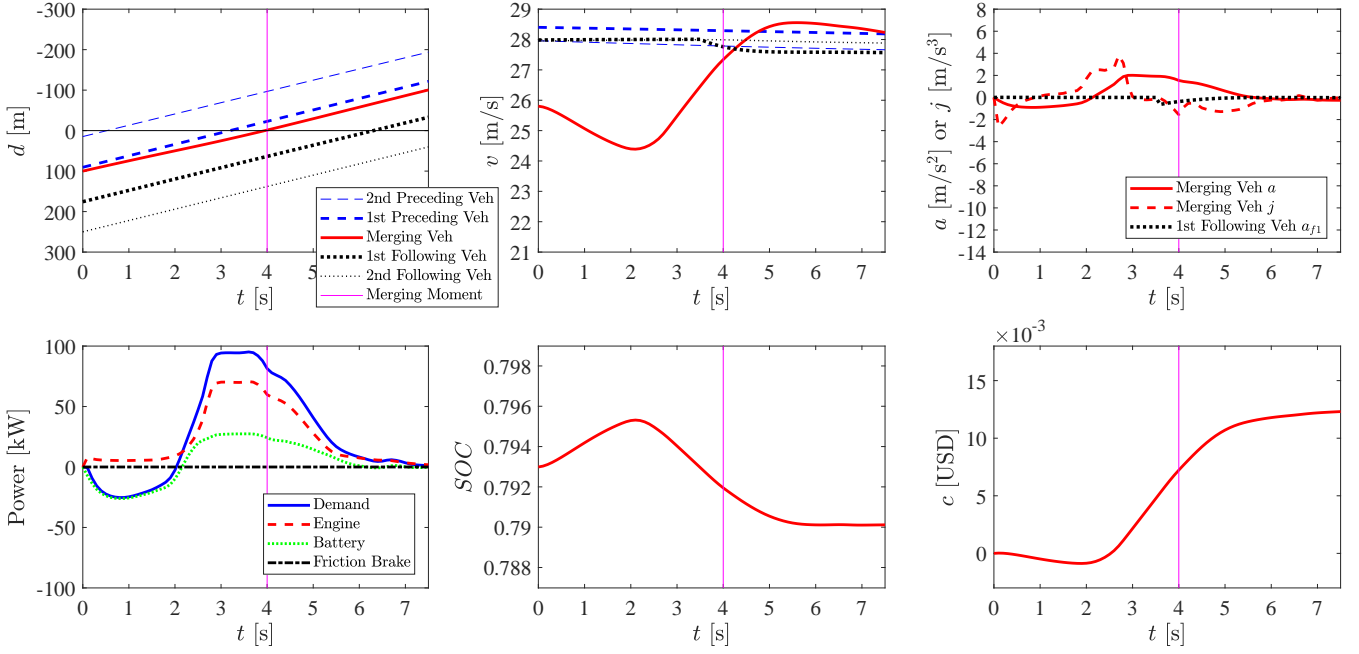}
\caption{A testing episode for co-optimization.}
\label{fig:episode_coop}
\end{figure*}

The testing episode for co-optimization is shown in Fig.~\ref{fig:episode_coop}. From the plot of distance to the merging point $d$, the merging vehicle merges ahead the first following vehicle. Additionally, the merging vehicle is closer to the first preceding vehicle than the first following vehicle. Even though one of the rewards dictates that the merging vehicle merges in the middle between two vehicles, the other reward that penalizes the braking of the first following vehicle dictates that the merging vehicle merges away from the first following vehicle. The combined effect results in the merging vehicle being closer to the first preceding vehicle in this episode.

From the plot of velocity $v$, the merging vehicle decelerates first, and then accelerates to reach the speed of the first preceding vehicle as dictated by one of the rewards.

The plot of acceleration $a$ and jerk $j$ shows that the merging vehicle has mild acceleration and jerk, indicating comfortable merging. The first following vehicle has very mild braking, indicating neglectable influence on the traffic flow. The first following vehicle's response to the merging vehicle is based on the IDM whose parameters are adjustable for different driving styles. Hence the braking of the first following vehicle is subjective evaluation.

The plot of the powertrain powers shows negative power demand during initial deceleration and positive power demand during acceleration. Friction brake power remains zero during the entire episode. During the initial deceleration, regenerative braking stores energy into the battery and hence negative battery power. The engine power is not zero during braking, possibly because SAC does not lead to zero action values. The non-zero engine power during braking results in more fuel consumption, indicating that the SAC policy is not optimal with regards to energy saving. During acceleration, the engine and battery co-power the vehicle.

The battery SOC increases during initial regenerative braking and decreases during acceleration. The instantaneous energy cost $c$ decreases during initial regenerative braking due to energy storage and increases as both the engine and battery power the vehicle.



\subsubsection{A testing episode for sequential approach II}

\begin{figure*}[htbp]
\centering
\includegraphics[width=7in]{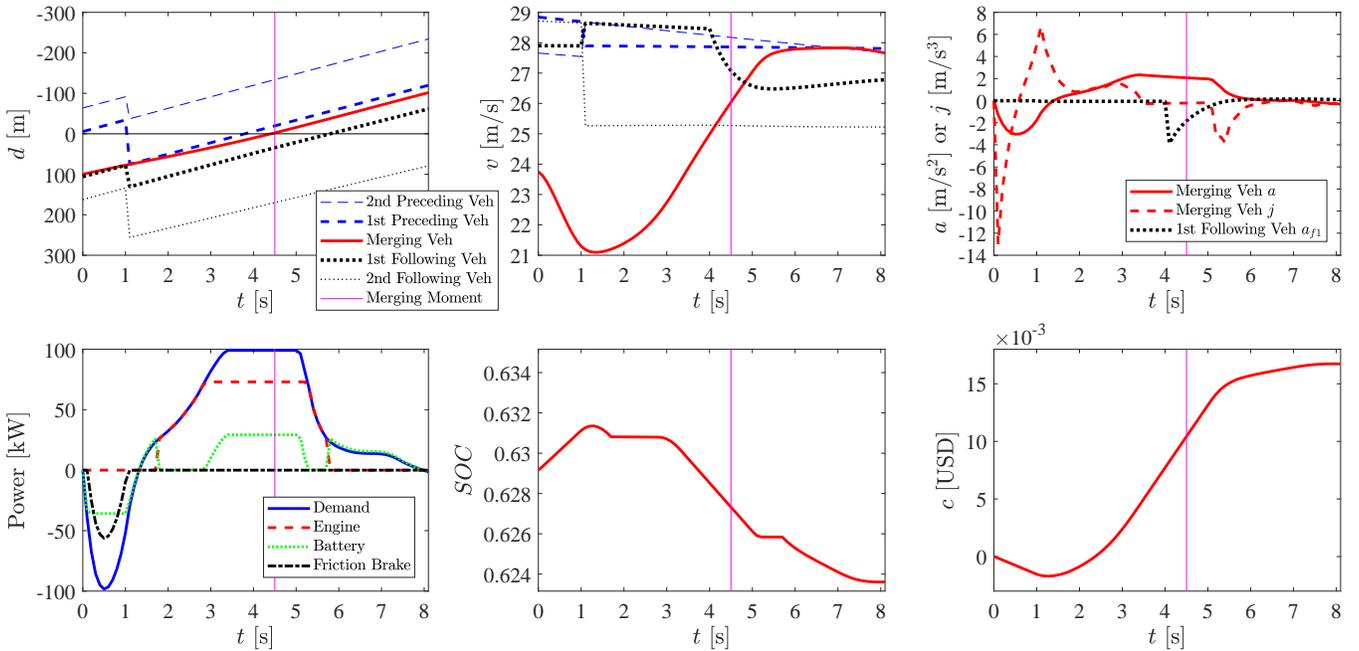}
\caption{A testing episode for sequential approach II wherein the upper-level DRL merging policy outputs the acceleration demand and the lower-level PHEV blended CD mode outputs the power split. $a$ denotes the actual acceleration that is different from the acceleration demand during power limit saturation. In the plot of distance to the merging point $d$, as the merging vehicle merges behind, the vehicles' positioning order suddenly changes at the 1s mark.}
\label{fig:episode_insuf}
\end{figure*}

Fig.~\ref{fig:episode_insuf} shows the testing episode for sequential approach II wherein the upper-level DRL merging policy outputs the acceleration demand. From the plot of distance to the merging point $d$, the merging vehicle merges behind. The merging vehicle's initial distance to the first following vehicle is relatively small, which may contribute to the decision of merging behind through learning.

From the plot of velocity $v$, the merging vehicle decelerates initially for the merging behind. Then it accelerates to match the speed of the first preceding vehicle as dictated by one of the rewards.

The merging vehicle's jerk $j$ has large absolute values initially for a short period of time, which results in passenger discomfort in this episode. Note that the issue of short-period large jerk exists for all three (co-optimization and two sequential) approaches.

From the plot of the powertrain powers, the vehicle brakes hard initially and regenerative braking is not able to provide all the braking power. Thus, friction braking is utilized as a supplement. There is power limit saturation of the power demand from 3.2s to 5s during acceleration, evidenced by both the engine and the battery reaching their power limits.

The battery SOC increases during initial regenerative braking and then decreases when the battery is utilized to power the vehicle. Correspondingly, the instantaneous energy cost decreases initially and increases when the battery and the engine co-power the vehicle.

\subsubsection{Testing results summary}

We define the following evaluation metrics to summarize the results of the co-optimization and the two sequential approaches.

(1) Power limit saturation rate. Power limit saturation rate is the number of episodes wherein the power demand is not achievable due to the powertrain power limits, divided by the total number of testing episodes. For sequential approach II, the acceleration demand can be transformed into power demand via the vehicle dynamics equation Eqn.~\ref{eq:Pd}.

(2) Collision rate. Collision rate is the number of episodes wherein the merging vehicle collides with another vehicle, divided by the total number of testing episodes.

(3) Average episodic energy cost in USD. Average episodic energy cost is the average of episodic energy costs over the total number of testing episodes. The energy cost is computed based on the instantaneous monetary energy cost definition in Eqn.~\ref{eq:cost}. The average episodic fuel and electricity costs are also computed separately to evaluate the power splits.

(4) Average jerk. Average jerk is defined as the average of the mean jerk of each episode over the total number of testing episodes.

(5) Merge-behind rate. Merge-behind rate is the number of episodes wherein the merging vehicle merges behind the first following vehicle, divided by the total number of testing episodes. Merging behind shows the decision-making of the merging vehicle.

A summary of the testing results is shown in Table~\ref{table:results}. Note that all the policies result in no stops.

\begin{table*}[!htbp]
\caption{Testing results summary. There are no stops for all approaches.}
\begin{center}
\begin{tabular}{|p{3cm}|p{1.5cm}|p{1.75cm}|p{1cm}|p{1cm}|p{1cm}|p{1.5cm}|p{1.5cm}|p{1.5cm}|}
\hline
Approach & \multirow{2}{1.5cm}{\# of testing episodes} & \multirow{2}{1.75cm}{Power limit saturation rate} & \multirow{2}{1cm}{Collision rate} & \multicolumn{3}{|l|}{Average episodic energy cost [USD]} & \multirow{2}{1.5cm}{Average jerk [m/s$^3$]} & \multirow{2}{1.5cm}{Merge-behind rate} \\
\cline{5-7}
 &  &  &  & fuel & electricity & combined &  &  \\
\hline
Co-optimization & 12831 & 0 & 0 & 0.013 & -0.002 & 0.011 (-8\%) & 1.0 (+11\%) & 7.5\% \\
\hline
Sequential approach I & 12668 & 0 & 0 & 0.010 & 0.002 & 0.012 (0\%) & 0.9 (0\%) & 8.7\% \\
\hline
Sequential approach II & 12835 & 25\% & 0.08\% & 0.009 & 0.002 & 0.011 (-8\%) & 0.8 (-11\%) & 6.9\% \\
\hline
\end{tabular}
\end{center}
\label{table:results}
\end{table*}

The power limit saturation rate is zero for co-optimization since co-optimization outputs the power split whose limits are inherently considered in the DRL framework. For sequential approach I wherein the upper-level merging policy outputs the power demand, the power limit saturation rate is also zero since the power limits are also inherently considered. For sequential approach II wherein the upper-level merging policy outputs the acceleration demand, the power demand may exceed the power limits as shown by the 25\% power limit saturation rate in Table~\ref{table:results}. When power limit saturation happens, the acceleration demand is not achieved, resulting in reduced actual acceleration magnitudes.

The collision rates are zero for co-optimization and sequential approach I. For sequential approach II, collisions happen at times due to power limit saturation and subsequent reduced merging performance. Note that when we test the DRL policy in another simulation where there are no power limits, the collision rate is zero for sequential approach II.

Sequential approach I has a higher energy cost than those of co-optimization and sequential approach II. This indicates that co-optimization finds economic power splits. However, the average episodic electricity cost for co-optimization is negative, suggesting that the trained co-optimization policy makes use of regenerative braking to harness energy.

The average jerk is less than or equal to 1m/s$^3$ for all cases, suggesting comfortable merging on average. However, the average jerk for co-optimization is higher than those for the two sequential approaches.

The merge-behind rates are only between 6.9\% to 8.7\% for all approaches. For majority of the episodes, the merging vehicle merges ahead the first following vehicle. Note that, in very limited episodes ($<0.01\%$), the merging vehicle merges ahead of the first preceding vehicle. This may in turn suggest the consideration of 2 vehicles ahead and behind in the DRL environment state for decision-making.

\section{Conclusions}

This work considers DRL-based on-ramp merging for a PHEV for the first time. We also consider DRL-based co-optimization for the first time and compare it with sequential approaches. In co-optimization, the on-ramp merging and the PHEV energy management are simultaneously optimized in a single DRL framework, directly outputting the PHEV power split for on-ramp merging. In sequential approaches, the upper-level DRL on-ramp merging policy outputs either the power or acceleration demand, which is utilized by the lower-level blended CD PHEV energy management for power split decisions.

The testing results show that the sequential approach II wherein the upper-level merging policy outputs the acceleration demand could lead to collisions if the acceleration demand cannot achieved due to PHEV power limit constraints. For co-optimization and the sequential approach I wherein the upper-level merging policy outputs the power demand, the power limits are inherently considered and the DRL-trained policies result in collision-free merging. This stresses the importance of considering power limit constraints in designing upper-level automated driving controllers.

Co-optimization offers comparable fuel economy compared to sequential approaches with the blended CD energy management. However, in co-optimization, the merging vehicle overwhelmingly performs regenerative braking for energy recuperation, suggesting that the merging behavior of co-optimization is different than that of the sequential approaches wherein the battery energy is depleted. We comment that since merging is a short-period event, energy saving for a single vehicle is secondary compared to safety.

The average jerk for co-optimization is higher than those for the sequential approaches. This is consistent with findings in the literature that co-optimization results in higher jerk. As suggested by the literature, the higher jerk could be a tradeoff with energy saving \cite{li2017fuel}.

The merging vehicle merges ahead in the majority of the episodes for the co-optimization and two sequential approaches, suggesting possibly aggressive merging. We believe that the decision to merge ahead or behind results from the designed rewards and the initial conditions.

Although we do not test the neural net solution on production vehicles, our previous work has suggested the real-time applicability of the same neural net size on automotive Electronic Control Units (ECUs). For the neural net size (2 hidden layers with 64 neurons each) considered, the execution time is an order of magnitude smaller than that of MPC with a linear model on a desktop computer \cite{lin2020comparison}. And MPC has been shown to be deployable in real time on automotive ECUs \cite{taghavipour2015real,vajedi2015ecological,batra2019real}. Thus, our neural net solution for on-ramp merging could satisfy the computational requirement for real-time deployment on production vehicles.

Future work includes testing the trained policies on real or scale vehicles \cite{lin2017}. We would also consider other DRL algorithms such as Twin-Delayed DDPG \cite{fujimoto2018addressing} for training.

\ifCLASSOPTIONcaptionsoff
  \newpage
\fi



%

\bibliographystyle{IEEEtran}
\bibliography{references}

\begin{thebibliography}{10}
\providecommand{\url}[1]{#1}
\csname url@samestyle\endcsname
\providecommand{\newblock}{\relax}
\providecommand{\bibinfo}[2]{#2}
\providecommand{\BIBentrySTDinterwordspacing}{\spaceskip=0pt\relax}
\providecommand{\BIBentryALTinterwordstretchfactor}{4}
\providecommand{\BIBentryALTinterwordspacing}{\spaceskip=\fontdimen2\font plus
\BIBentryALTinterwordstretchfactor\fontdimen3\font minus
  \fontdimen4\font\relax}
\providecommand{\BIBforeignlanguage}[2]{{%
\expandafter\ifx\csname l@#1\endcsname\relax
\typeout{** WARNING: IEEEtran.bst: No hyphenation pattern has been}%
\typeout{** loaded for the language `#1'. Using the pattern for}%
\typeout{** the default language instead.}%
\else
\language=\csname l@#1\endcsname
\fi
#2}}
\providecommand{\BIBdecl}{\relax}
\BIBdecl

\bibitem{rios2016survey}
J.~Rios-Torres and A.~A. Malikopoulos, ``A survey on the coordination of
  connected and automated vehicles at intersections and merging at highway
  on-ramps,'' \emph{IEEE Transactions on Intelligent Transportation Systems},
  vol.~18, no.~5, pp. 1066--1077, 2016.

\bibitem{treiber2000congested}
M.~Treiber, A.~Hennecke, and D.~Helbing, ``Congested traffic states in
  empirical observations and microscopic simulations,'' \emph{Physical Review
  E}, vol.~62, no.~2, p. 1805, 2000.

\bibitem{rios2017automated}
J.~Rios-Torres and A.~A. Malikopoulos, ``Automated and cooperative vehicle
  merging at highway on-ramps,'' \emph{IEEE Transactions on Intelligent
  Transportation Systems}, vol.~18, no.~4, pp. 780--789, 2017.

\bibitem{tran2019model}
A.~T. Tran, M.~Kawaguchi, H.~Okuda, and T.~Suzuki, ``A model predictive
  control-based lane merging strategy for autonomous vehicles,'' in \emph{2019
  IEEE Intelligent Vehicles Symposium (IV)}.\hskip 1em plus 0.5em minus
  0.4em\relax IEEE, 2019, pp. 594--599.

\bibitem{nishi2019merging}
T.~Nishi, P.~Doshi, and D.~Prokhorov, ``Merging in congested freeway traffic
  using multipolicy decision making and passive actor-critic learning,''
  \emph{IEEE Transactions on Intelligent Vehicles}, vol.~4, no.~2, pp.
  287--297, 2019.

\bibitem{sutton2018reinforcement}
R.~S. Sutton and A.~G. Barto, \emph{Reinforcement learning: An
  introduction}.\hskip 1em plus 0.5em minus 0.4em\relax MIT press, 2018.

\bibitem{silver2016mastering}
D.~Silver, A.~Huang, C.~J. Maddison, A.~Guez, L.~Sifre, G.~Van Den~Driessche,
  J.~Schrittwieser, I.~Antonoglou, V.~Panneershelvam, M.~Lanctot \emph{et~al.},
  ``Mastering the game of go with deep neural networks and tree search,''
  \emph{Nature}, vol. 529, no. 7587, pp. 484--489, 2016.

\bibitem{lin2020comparison}
Y.~Lin, J.~McPhee, and N.~L. Azad, ``Comparison of deep reinforcement learning
  and model predictive control for adaptive cruise control,'' \emph{IEEE
  Transactions on Intelligent Vehicles}, vol.~6, no.~2, pp. 221--231, 2021.

\bibitem{hu2019interaction}
Y.~Hu, A.~Nakhaei, M.~Tomizuka, and K.~Fujimura, ``Interaction-aware decision
  making with adaptive strategies under merging scenarios,'' in \emph{2019
  IEEE/RSJ International Conference on Intelligent Robots and Systems
  (IROS)}.\hskip 1em plus 0.5em minus 0.4em\relax IEEE, 2019, pp. 151--158.

\bibitem{bouton2019cooperation}
M.~Bouton, A.~Nakhaei, K.~Fujimura, and M.~J. Kochenderfer, ``Cooperation-aware
  reinforcement learning for merging in dense traffic,'' in \emph{2019 IEEE
  Intelligent Transportation Systems Conference (ITSC)}.\hskip 1em plus 0.5em
  minus 0.4em\relax IEEE, 2019, pp. 3441--3447.

\bibitem{lin2020anti}
Y.~Lin, J.~McPhee, and N.~L. Azad, ``Anti-jerk on-ramp merging using deep
  reinforcement learning,'' in \emph{2020 IEEE Intelligent Vehicles Symposium
  (IV)}.\hskip 1em plus 0.5em minus 0.4em\relax IEEE, 2020, pp. 7--14.

\bibitem{liu2008modeling}
J.~Liu and H.~Peng, ``Modeling and control of a power-split hybrid vehicle,''
  \emph{IEEE Transactions on Control Systems Technology}, vol.~16, no.~6, pp.
  1242--1251, 2008.

\bibitem{gao1999investigation}
Y.~Gao, L.~Chen, and M.~Ehsani, ``Investigation of the effectiveness of
  regenerative braking for {EV} and {HEV},'' in \emph{Future Transportation
  Technology Conference \& Exposition}.\hskip 1em plus 0.5em minus 0.4em\relax
  SAE International, 1999.

\bibitem{shao2019optimal}
Y.~Shao and Z.~Sun, ``Optimal speed control for a connected and autonomous
  electric vehicle considering battery aging and regenerative braking limits,''
  in \emph{Dynamic Systems and Control Conference}, vol. 59148.\hskip 1em plus
  0.5em minus 0.4em\relax American Society of Mechanical Engineers, 2019, p.
  V001T08A005.

\bibitem{taghavipour2019intelligent}
A.~Taghavipour, M.~Vajedi, and N.~L. Azad, \emph{Intelligent control of
  connected plug-in hybrid electric vehicles}.\hskip 1em plus 0.5em minus
  0.4em\relax Springer, 2019.

\bibitem{jeong2019analysis}
J.~Jeong, N.~Kim, K.~Stutenberg, and A.~Rousseau, ``Analysis and model
  validation of the {T}oyota {P}rius {P}rime,'' in \emph{WCX SAE World Congress
  Experience}.\hskip 1em plus 0.5em minus 0.4em\relax SAE International, 2019.

\bibitem{taghavipour2015real}
A.~Taghavipour, N.~L. Azad, and J.~McPhee, ``Real-time predictive control
  strategy for a plug-in hybrid electric powertrain,'' \emph{Mechatronics},
  vol.~29, pp. 13--27, 2015.

\bibitem{taghavipour2016comparative}
A.~Taghavipour, M.~Vajedi, N.~L. Azad, and J.~McPhee, ``A comparative analysis
  of route-based energy management systems for {PHEV}s,'' \emph{Asian Journal
  of Control}, vol.~18, no.~1, pp. 29--39, 2016.

\bibitem{lin2003power}
C.-C. Lin, H.~Peng, J.~W. Grizzle, and J.-M. Kang, ``Power management strategy
  for a parallel hybrid electric truck,'' \emph{IEEE Transactions on Control
  Systems Technology}, vol.~11, no.~6, pp. 839--849, 2003.

\bibitem{banvait2009rule}
H.~Banvait, S.~Anwar, and Y.~Chen, ``A rule-based energy management strategy
  for plug-in hybrid electric vehicle ({PHEV}),'' in \emph{2009 American
  Control Conference}.\hskip 1em plus 0.5em minus 0.4em\relax IEEE, 2009, pp.
  3938--3943.

\bibitem{kim2010optimal}
N.~Kim, S.~Cha, and H.~Peng, ``Optimal control of hybrid electric vehicles
  based on {P}ontryagin's minimum principle,'' \emph{IEEE Transactions on
  Control Systems Technology}, vol.~19, no.~5, pp. 1279--1287, 2010.

\bibitem{musardo2005ecms}
C.~Musardo, G.~Rizzoni, Y.~Guezennec, and B.~Staccia, ``{A-ECMS}: An adaptive
  algorithm for hybrid electric vehicle energy management,'' \emph{European
  Journal of Control}, vol.~11, no. 4-5, pp. 509--524, 2005.

\bibitem{borhan2011mpc}
H.~Borhan, A.~Vahidi, A.~M. Phillips, M.~L. Kuang, I.~V. Kolmanovsky, and
  S.~Di~Cairano, ``{MPC}-based energy management of a power-split hybrid
  electric vehicle,'' \emph{IEEE Transactions on Control Systems Technology},
  vol.~20, no.~3, pp. 593--603, 2011.

\bibitem{serrao2011comparative}
L.~Serrao, S.~Onori, and G.~Rizzoni, ``A comparative analysis of energy
  management strategies for hybrid electric vehicles,'' \emph{Journal of
  Dynamic Systems, Measurement, and Control}, vol. 133, no.~3, 2011.

\bibitem{lin2014reinforcement}
X.~Lin, Y.~Wang, P.~Bogdan, N.~Chang, and M.~Pedram, ``Reinforcement learning
  based power management for hybrid electric vehicles,'' in \emph{2014 IEEE/ACM
  International Conference on Computer-Aided Design (ICCAD)}.\hskip 1em plus
  0.5em minus 0.4em\relax IEEE, 2014, pp. 33--38.

\bibitem{liu2014power}
C.~Liu and Y.~L. Murphey, ``Power management for plug-in hybrid electric
  vehicles using reinforcement learning with trip information,'' in \emph{2014
  IEEE Transportation Electrification Conference and Expo (ITEC)}.\hskip 1em
  plus 0.5em minus 0.4em\relax IEEE, 2014, pp. 1--6.

\bibitem{li2019deep}
Y.~Li, H.~He, J.~Peng, and H.~Wang, ``Deep reinforcement learning-based energy
  management for a series hybrid electric vehicle enabled by history cumulative
  trip information,'' \emph{IEEE Transactions on Vehicular Technology},
  vol.~68, no.~8, pp. 7416--7430, 2019.

\bibitem{xu2020ensemble}
B.~Xu, X.~Hu, X.~Tang, X.~Lin, H.~Li, D.~Rathod, and Z.~Filipi, ``Ensemble
  reinforcement learning-based supervisory control of hybrid electric vehicle
  for fuel economy improvement,'' \emph{IEEE Transactions on Transportation
  Electrification}, vol.~6, no.~2, pp. 717--727, 2020.

\bibitem{qi2019deep}
X.~Qi, Y.~Luo, G.~Wu, K.~Boriboonsomsin, and M.~Barth, ``Deep reinforcement
  learning enabled self-learning control for energy efficient driving,''
  \emph{Transportation Research Part C: Emerging Technologies}, vol.~99, pp.
  67--81, 2019.

\bibitem{liu2019heuristic}
T.~Liu, X.~Hu, W.~Hu, and Y.~Zou, ``A heuristic planning reinforcement
  learning-based energy management for power-split plug-in hybrid electric
  vehicles,'' \emph{IEEE Transactions on Industrial Informatics}, vol.~15,
  no.~12, pp. 6436--6445, 2019.

\bibitem{liu2017reinforcement}
T.~Liu, X.~Hu, S.~E. Li, and D.~Cao, ``Reinforcement learning optimized
  look-ahead energy management of a parallel hybrid electric vehicle,''
  \emph{IEEE/ASME Transactions on Mechatronics}, vol.~22, no.~4, pp.
  1497--1507, 2017.

\bibitem{zhu2020energy}
Z.~Zhu, Y.~Liu, and M.~Canova, ``Energy management of hybrid electric vehicles
  via deep {Q}-networks,'' in \emph{2020 American Control Conference}.\hskip
  1em plus 0.5em minus 0.4em\relax IEEE, 2020, pp. 3077--3082.

\bibitem{chen2019series}
B.~Chen, S.~A. Evangelou, and R.~Lot, ``Series hybrid electric vehicle
  simultaneous energy management and driving speed optimization,''
  \emph{IEEE/ASME Transactions on Mechatronics}, vol.~24, no.~6, pp.
  2756--2767, 2019.

\bibitem{chen2020iterative}
D.~Chen, Y.~Kim, M.~Huang, and A.~Stefanopoulou, ``An iterative and
  hierarchical approach to co-optimizing the velocity profile and power-split
  of plug-in hybrid electric vehicles,'' in \emph{2020 American Control
  Conference (ACC)}.\hskip 1em plus 0.5em minus 0.4em\relax IEEE, 2020, pp.
  3059--3064.

\bibitem{li2017fuel}
L.~Li, X.~Wang, and J.~Song, ``Fuel consumption optimization for smart hybrid
  electric vehicle during a car-following process,'' \emph{Mechanical Systems
  and Signal Processing}, vol.~87, pp. 17--29, 2017.

\bibitem{xu2021decentralized}
F.~Xu and T.~Shen, ``Decentralized optimal merging control with optimization of
  energy consumption for connected hybrid electric vehicles,'' \emph{IEEE
  Transactions on Intelligent Transportation Systems}, 2021.

\bibitem{vajedi2015ecological}
M.~Vajedi and N.~L. Azad, ``Ecological adaptive cruise controller for plug-in
  hybrid electric vehicles using nonlinear model predictive control,''
  \emph{IEEE Transactions on Intelligent Transportation Systems}, vol.~17,
  no.~1, pp. 113--122, 2015.

\bibitem{taghavipour2013high}
A.~Taghavipour, R.~Masoudi, N.~L.~Azad, and J.~McPhee, ``High-fidelity modeling
  of a power-split plug-in hybrid electric powertrain for control performance
  evaluation,'' in \emph{International Design Engineering Technical Conferences
  and Computers and Information in Engineering Conference}, vol. 55843.\hskip
  1em plus 0.5em minus 0.4em\relax American Society of Mechanical Engineers,
  2013, p. V001T01A008.

\bibitem{buggaveeti2017longitudinal}
S.~Buggaveeti, M.~Batra, J.~McPhee, and N.~Azad, ``Longitudinal vehicle
  dynamics modeling and parameter estimation for plug-in hybrid electric
  vehicle,'' \emph{SAE International Journal of Vehicle Dynamics, Stability,
  and NVH}, vol.~1, no. 2017-01-1574, pp. 289--297, 2017.

\bibitem{fersch2017challenges}
T.~Fersch, R.~Weigel, and A.~Koelpin, ``Challenges in miniaturized automotive
  long-range lidar system design,'' in \emph{Three-Dimensional Imaging,
  Visualization, and Display 2017}, vol. 10219.\hskip 1em plus 0.5em minus
  0.4em\relax International Society for Optics and Photonics, 2017, p. 102190T.

\bibitem{hecht2018lidar}
J.~Hecht, ``Lidar for self-driving cars,'' \emph{Optics and Photonics News},
  vol.~29, no.~1, pp. 26--33, 2018.

\bibitem{aashto2001policy}
A.~Aashto, ``Policy on geometric design of highways and streets,''
  \emph{American Association of State Highway and Transportation Officials,
  Washington, DC}, vol.~1, no. 990, p. 158, 2001.

\bibitem{batra2019real}
M.~Batra, J.~McPhee, and N.~L. Azad, ``Real-time model predictive control of
  connected electric vehicles,'' \emph{Vehicle System Dynamics}, vol.~57,
  no.~11, pp. 1720--1743, 2019.

\bibitem{haarnoja2018soft}
T.~Haarnoja, A.~Zhou, P.~Abbeel, and S.~Levine, ``Soft actor-critic: Off-policy
  maximum entropy deep reinforcement learning with a stochastic actor,'' in
  \emph{International Conference on Machine Learning}, 2018, pp. 1861--1870.

\bibitem{ioffe2015batch}
S.~Ioffe and C.~Szegedy, ``Batch normalization: Accelerating deep network
  training by reducing internal covariate shift,'' in \emph{International
  conference on machine learning}.\hskip 1em plus 0.5em minus 0.4em\relax PMLR,
  2015, pp. 448--456.

\bibitem{van2016deep}
H.~Van~Hasselt, A.~Guez, and D.~Silver, ``Deep reinforcement learning with
  double {Q}-learning,'' in \emph{Proceedings of the AAAI conference on
  artificial intelligence}, vol.~30, no.~1, 2016.

\bibitem{lillicrap2015continuous}
T.~P. Lillicrap, J.~J. Hunt, A.~Pritzel, N.~Heess, T.~Erez, Y.~Tassa,
  D.~Silver, and D.~Wierstra, ``Continuous control with deep reinforcement
  learning,'' \emph{arXiv preprint arXiv:1509.02971}, 2015.

\bibitem{schulman2017proximal}
J.~Schulman, F.~Wolski, P.~Dhariwal, A.~Radford, and O.~Klimov, ``Proximal
  policy optimization algorithms,'' \emph{arXiv preprint arXiv:1707.06347},
  2017.

\bibitem{lin2017}
Y.~Lin and A.~Eskandarian, ``Experimental evaluation of cooperative adaptive
  cruise control with autonomous mobile robots,'' in \emph{2017 IEEE Conference
  on Control Technology and Applications (CCTA)}.\hskip 1em plus 0.5em minus
  0.4em\relax IEEE, 2017, pp. 281--286.

\bibitem{fujimoto2018addressing}
S.~Fujimoto, H.~Hoof, and D.~Meger, ``Addressing function approximation error
  in actor-critic methods,'' in \emph{International Conference on Machine
  Learning}, 2018, pp. 1587--1596.

\end{thebibliography}




%

\begin{IEEEbiography}[{\includegraphics[width=1in,height=1.25in,clip,keepaspectratio]{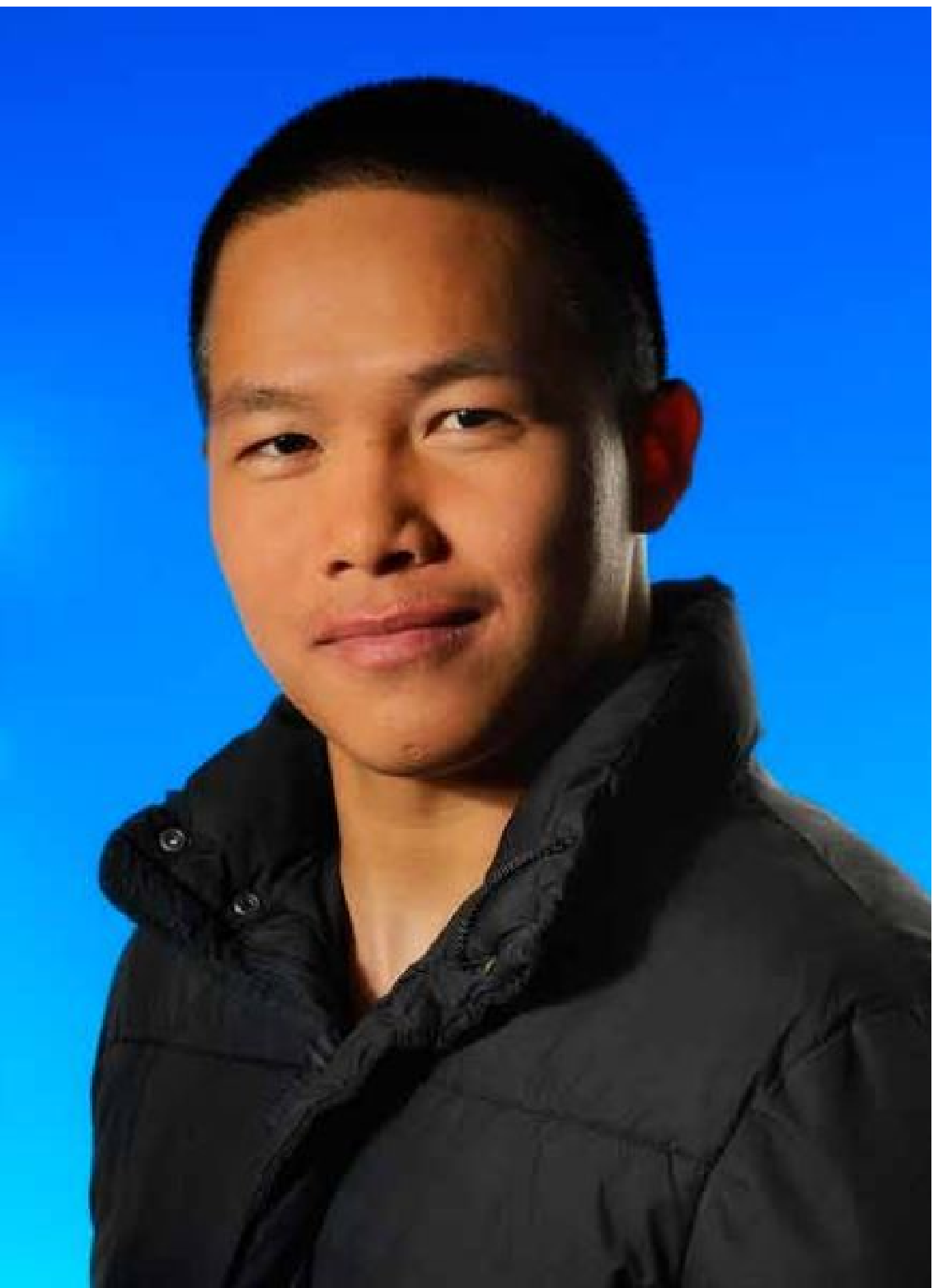}}]{Yuan Lin} received the B.E. degree in Civil Engineering from Nanchang University, China, in 2011 and the Ph.D. degree in Engineering Mechanics from Virginia Tech, Blacksburg, VA, USA, in 2016. He was a Postdoctoral Fellow in the Mechanical Engineering Department at Virginia Tech from 2016 to 2018 and in the Systems Design Engineering Department at the University of Waterloo from 2018 to 2020. He is currently an assistant professor in the Shien-Ming Wu School of Intelligent Engineering at the South China University of Technology. His research interests include autonomous driving, reinforcement learning, and hybrid electric vehicles. \end{IEEEbiography}

\begin{IEEEbiography}[{\includegraphics[width=1in,height=1.25in,clip,keepaspectratio]{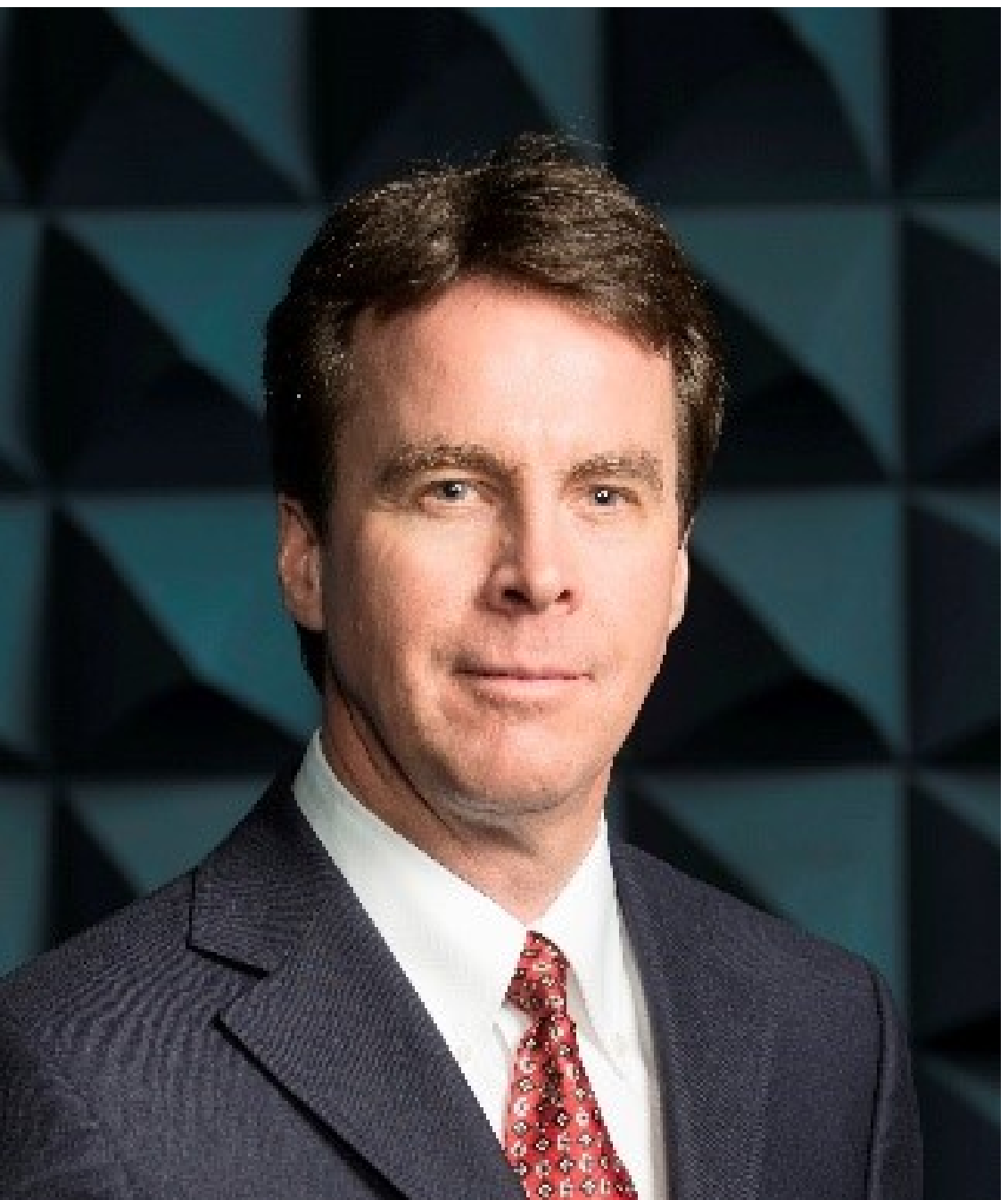}}]{Professor John McPhee} is the Canada Research Chair in System Dynamics at the University of Waterloo, Canada, which he joined in 1992.  Prior to that, he held fellowships at Queen’s University, Canada, and the Universite de Liege, Belgium.
He pioneered the use of linear graph theory and symbolic computing to create real-time models and model-based controllers for multi-domain dynamic systems, with applications ranging from autonomous vehicles to rehabilitation robots and sports engineering. His research algorithms are a core component of the widely-used MapleSim modelling software, and his work appears in more than 160 journal publications.
Prof. McPhee is the past Chair of the International Association for Multibody System Dynamics, a co-founder of 2 international journals and 3 technical committees, a member of the Golf Digest Technical Panel, and an Associate Editor for 5 journals. He is a Fellow of the Canadian Academy of Engineering, the American and Canadian Societies of Mechanical Engineers, and the Engineering Institute of Canada.  He has won 8 Best Paper Awards and, in 2014, he received the prestigious NSERC Synergy Award from the Governor-General of Canada. \end{IEEEbiography}

\begin{IEEEbiography}[{\includegraphics[width=1in,height=1.25in,clip,keepaspectratio]{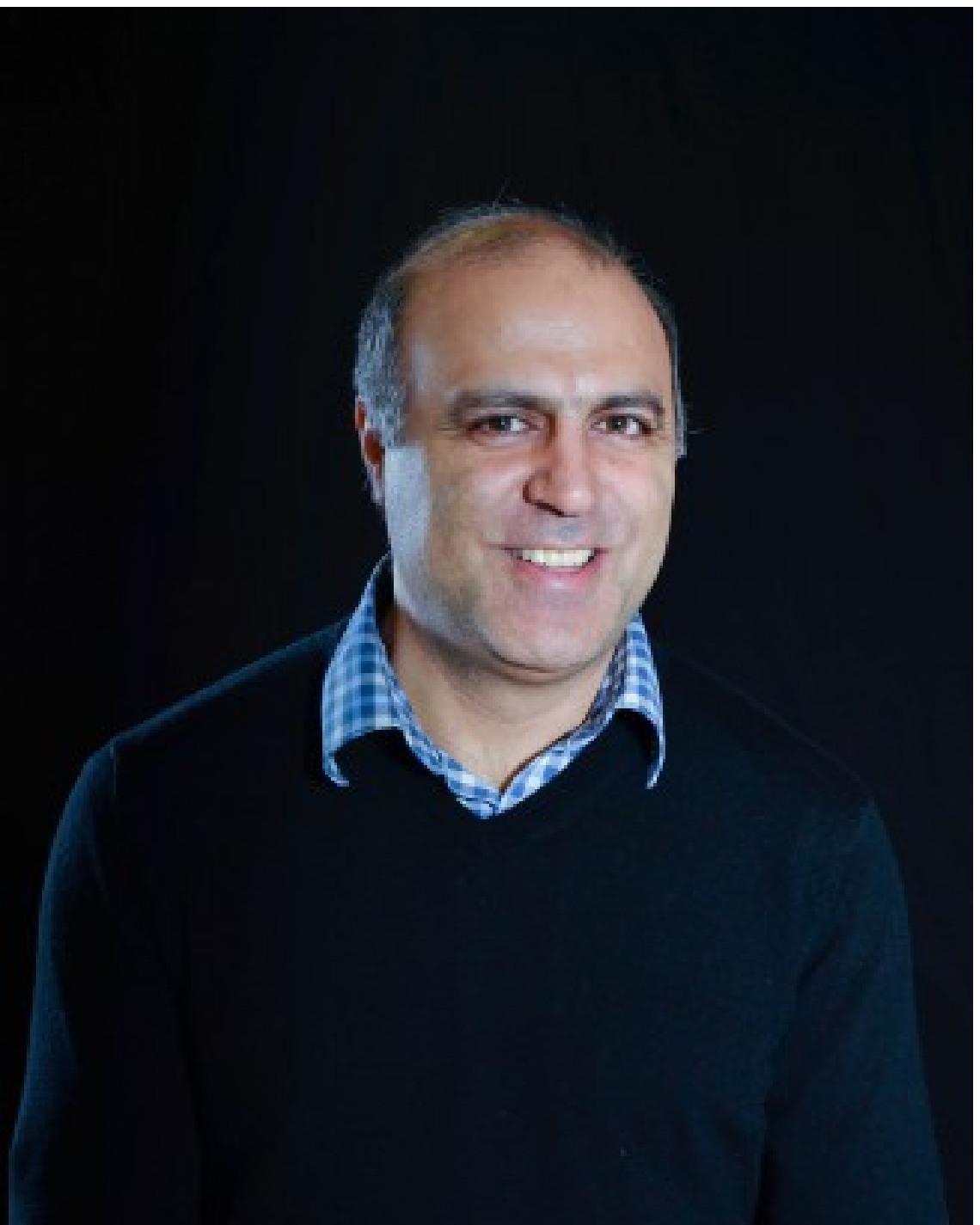}}]{Nasser L. Azad} is currently an Associate Professor in the Department of Systems Design Engineering, University of Waterloo, and the Director of Smart Hybrid and Electric Vehicle Systems (SHEVS) Laboratory. He was a Postdoctoral Fellow with the Vehicle Dynamics and Control Laboratory, Department of Mechanical Engineering, University of California, Berkeley, CA, USA. Dr. Azad’s primary research interests lie in control of connected hybrid and electric vehicles, autonomous cars, and unmanned aerial vehicle quad-rotors. He is also interested in applications of Artificial Intelligence for solving different engineering problems. Due to his outstanding work, Dr. Azad received an Early Researcher Award in 2015 from the Ministry of Research and Innovation, Ontario, Canada. \end{IEEEbiography}

%
%




\end{document}